\documentclass[showpacs,preprintnumbers,superscriptaddress,amsmath,amssymb,floatfix]{revtex4}
\usepackage{amsmath}
\usepackage{mathpazo}
\usepackage{epsfig}
\usepackage{graphics}
\usepackage{color}
\usepackage{graphicx}
\usepackage[font={footnotesize,it}]{caption}
\usepackage[colorlinks=true,
            linkcolor=black,
            urlcolor=blue,
            citecolor=blue]{hyperref}
\begin{document}

\title{Deflection of light by black holes and massless wormholes in massive gravity }

\author{Kimet Jusufi}
\email{kimet.jusufi@unite.edu.mk}
\affiliation{Physics Department, State University of Tetovo, Ilinden Street nn, 1200,
Tetovo, Macedonia}
\affiliation{Institute of Physics, Faculty of Natural Sciences and Mathematics, Ss. Cyril and Methodius University, Arhimedova 3, 1000 Skopje, Macedonia}

\author{Nayan Sarkar}
 \email{nayan.mathju@gmail.com}
\affiliation {Department of Mathematics, Jadavpur University,
 Kolkata-700032, India.}

\author{Farook Rahaman}
 \email{rahaman@associates.iucaa.in}
\affiliation {Department of Mathematics, Jadavpur University,
 Kolkata-700032, India.}

\author{Ayan Banerjee}
 \email{ayan_7575@yahoo.co.in}
 \affiliation{Astrophysics and Cosmology Research Unit, University of KwaZulu Natal, Private Bag X54001, Durban 4000,
South Africa.}
\affiliation {Department of Mathematics, Jadavpur University,
 Kolkata-700032, India.}

 \author{Sudan Hansraj}
 \email{hansrajs@ukzn.ac.za}
 \affiliation{Astrophysics and Cosmology Research Unit, University of KwaZulu Natal, Private Bag X54001, Durban 4000,
South Africa.}

\date{\today }

\begin{abstract}
Weak gravitational lensing by black holes and wormholes in the context of massive gravity (Bebronne and Tinyakov 2009) theory is studied.  The particular solution examined is characterized by two integration constants, the mass $M$ and an extra parameter $S$ namely `scalar charge'. These black hole reduce to the standard Schwarzschild black hole solutions when the scalar charge is zero and the mass is positive. In addition, a parameter $\lambda$ in the metric characterizes so-called 'hair'. The geodesic equations are used to examine the behavior of the deflection angle in four relevant cases of the parameter $\lambda$. Then, by introducing a simple coordinate transformation $r^\lambda=S+v^2$ into the black hole metric, we were able to find a massless wormhole solution of Einstein-Rosen (ER) \cite{Einstein} type with scalar charge $S$. The programme is then repeated in terms of the Gauss--Bonnet theorem in the weak field limit after a method is established to deal with the angle of deflection using different domains of integration depending on the parameter $\lambda$.  In particular, we have found new analytical results corresponding to  four special cases which generalize the well known deflection angles reported in the literature. Finally, we have established the time delay problem in the spacetime of black holes and wormholes, respectively.            
\end{abstract}

\maketitle

\section{Introduction} 
At  present independent observations have confirmed that the universe is currently undergoing a phase of accelerated expansion. The observed late time acceleration has been confirmed by  data from type Ia Supernovae \cite{Riess}, anisotropy in the Cosmic Microwave Background radiation \cite{Spergel} and SDSS \cite{Tegmark,Tegmark1}. To describe the present expansion scenario several models have been proposed so far.  Two broad approaches have emerged to account for the observed accelerated expansion.  The first is the dark energy proposal with the assumption that  nearly 70 \% of the total energy-density in the universe may be in the form of negative pressure fluid with the associated density parameter $\Omega_{\text{DE}}$ of the order of $\Omega_{\text{DE}}$ $\sim$ 0.70. One of the simplest candidates generating the  dark energy is the cosmological constant, but its characterization has two well-known problems, i.e., fine-tuning and cosmic coincidence. Moreover, there is a severe discrepancy in the observed value of the cosmological constant in contrast with the value predicted by quantum cosmology. Ellis {\it{et al}} \cite{ellis1,ellis2} proposed the use of the trace-free Einstein equations which effectively treats the cosmological constant as a mere constant of integration. This idea was first proposed by Weinberg \cite{weinberg2} and has also gone by the name unimodular gravity \cite{unimod1,unimod2,unimod3}.    Several alternative models have been suggested to incorporate the cosmological constant problems, namely, quintessence \cite{Caldwell}, tachyon field \cite{Gorini}, phantom model \cite{Carroll} and $k$-essence \citep{Chimento} that also predict cosmic expansion amongst others.

 A second approach is that of modified gravity as an alternative to  appealing to exotic matter distributions such as dark energy or dark matter. Generalizations of general relativity (GR) appear  to avoid introducing matter with nonstandard physical properties and to solve the singularity problem. Modified or extended theories of gravity often require higher dimensional spacetimes. This in itself is no shortcoming as historically a number of higher dimensional theories have appeared such as Kaluza-Klein theory and the brane world concept. It is debatable whether gravitational interactions are necessarily four dimensional. Indeed if string theory or its generalization M-theory for quantum effects is to be consistent with a theory of gravitation then higher dimensions are necessary. The Einstein-Hilbert action may be modified to  include non-linear geometric terms. One of these proposals is the $f(R)$ theory \citep{Nojiri,starobinksi}, as a simple modification of the Einstein-Hilbert Lagrangian density by a general function of the Ricci scalar $R$. While $f(R)$ theory does have the capacity to explain the late-time expansion of the universe, the theory does possess some difficulties in that ghost terms are manifest in the presence of fourth order derivatives. Of late $f(R)$ theory has been shown to be equivalent to the Brans-Dicke scalar tensor theory.  A more natural generalization of general relativity is the Lovelock \cite{love1,love2} lagrangian postulate in which the action is composed of terms quadratic in the Ricci scalar, Ricci tensor and the Riemann tensor. Remarkably this higher curvature theory generates up to second order derivative terms in the equations of motion and is accordingly ghost-free. To zeroth order the Lovelock polynomial is identical to the cosmological constant, to first order the Einstein action is regained while to second order the action is known as the Gauss--Bonnet action.

In this paper, we consider massive gravity  as a modification of GR. These  include massive gravitons and have attracted much attention recently. In addition the theory incorporates massive spin-2 particles which have two degrees of freedom. This theory has a rich phenomenology, such as explaining the accelerated expansion of the universe without invoking dark energy. Additionally,  the resolution of the hierarchy problem and  brane-world gravity scenarios also generate arguments for  the existence of massive modes; hence massive gravity as in the Ref. \cite{Dvali,Dvali1} emerged. In this direction the pioneer work was done by Fierz and Pauli \cite{Fierz} in the context of linear theory.
It is worthwhile to mention that the original theory suffered from the existence of vDVZ (van Dam-Veltman-Zakharov) discontinuity. Later, Vainshtein introduced a well known mechanism \cite{Vainshtein} to resolve the long standing problem of the vDVZ discontinuity by considering  a nonlinear framework but this raised another problem of the Fierz and Pauli theory which is
known as the Boulware-Deser (BD) \cite{Boulware} ghost instability at the non-linear level. In order to avoid such instability, de Rham, Gabadadze and Tolley (dRGT) \cite{Rham}  have proposed a new massive gravity theory  with an extension of the Fierz-Pauli theory. Recently other versions of  massive gravity have been proposed, namely, new massive gravity \cite{Bergshoeff} and bi-gravity \cite{Hassan}.

Massive gravity theories are also studied in the astrophysical context.  Black hole solutions and their thermodynamical properties have been analyzed in dRGT massive gravity \cite{Cai}. Katsuragawa \textit{et al.}  \cite{Katsuragawa} devised  a neutron star model  that demonstrated  that   massive gravity dynamics deviates only slightly from  GR. It was recently proposed by Bebronne and Tinyakov \cite{bebronne} that  vacuum spherically-symmetric solutions  do exist in massive gravity. The black hole solution depends on the mass  $M$ and an extra parameter $S$ which is referred to as  the `scalar charge'. Additionally, in Ref. \cite{Capela}  the validity of the laws of thermodynamics in massive gravity have been checked for the same black hole solutions. A number of articles on  black holes in massive gravity have appeared recently; some solutions have been reported in \cite{Xu}.

It is important to understand the deflection of light in the presence of a mass distribution. This becomes an important and effective tool for probing a number of interesting phenomena.  As early as  1919 Eddington  \cite{Eddington} studied the  weak gravitational lensing of the Schwarzschild spacetime. This seminal work initiated the study of gravitational lensing (GL) theory \cite{Schneider}. It is also known that in the vicinity of massive compact objects (such as neutron stars or black holes) electromagnetic radiation is generated. The importance of examining light deflection in the weak field limit lies in  the ability to probe  large-scale structures, as well as exotic matter, wormholes, naked singularity, etc (The reader is referred to the more detailed review in \cite{Saravani}). It is thus imperative to investigate the GL effect  of black holes in massive gravity and to search for their possible observational signatures in the weak field limit.  In contrast to the lensing situation already studied in the literature, we apply  the higher curvature Gauss-Bonnet theorem (GTB) \cite{gibbons1} to calculate the deflection angle.

It is well known that the deflection of light (i.e. Gravitational lensing) is now one of useful tools to search not only for dark and massive objects, but also wormholes. In recently past, several attempts have been made to calculate the elliptical integral by Virbhadra and Ellis \cite{Virbhadra}. Soon after the Eiroa et al. have studited Riessner-Nordstrom black hole lensing in strong gravitational region \cite{Eiroa}. The  black hole gravitational lenses have been widely demonstrated in \cite{Ghaffarnejad}. In addition, after the pioneer works by Kim and Cho \cite{Kim}, the gravitational lensing by a negative Arnowitt-Deser-Misner (ADM) mass was studied in \cite{Cramer}. As a consequence, several forms of the deflection angle by the Ellis wormhole (particular example of the Morris-Thorne traversable wormhole) have been studied in the strong field limit \cite{Nandi}. The  computation of the  deflection angle in the weak field limit  for spherically symmetric static spacetimes may be accomplished through a simple algorithm. Very recently, Werner \cite{Werner} extended and applied the optical geometry  to the case of  stationary black holes. Further, under some physically realistic assumptions GBT was used in  studies of various astrophysical objects, such as Ellis wormholes by Jusufi \cite{kimet1}, wormholes in Einstein-Maxwell-dilaton theory \cite{goulart},  black holes with topological defects and deflection angle for finite distance by Ishihara et al. \cite{kimet1,ishihara}. In Ref. \cite{zhang}, the authors have studied the strong deflection limit from black holes and explored the role of the  scalar charge in massive gravity.  In the present work, we aim to investigate the deflection angle by black holes and charged wormholes in massive gravity in the weak limit approximation using the optical geometry as well as the geodesic method.\\

This paper is  structured as follows. In Sec. \textbf{II} we review  the black hole solution in massive gravity. In Sec. \textbf{III} we  consider the geodesic equations in massive gravity theory and analyse the deflection angle in four special cases. In Sec. IV we consider the same problem viewed in terms of the Gauss-Bonnet theorem. In Sec. \textbf{V} the time delay problem is considered. In Sec. \textbf{VI } we shall consider deflection of light by wormholes. By applying the GBT of gravitational lensing theory to the optical geometry, we calculate the deflection angle produced by charged and massless wormhole in massive gravity. In Sec. \textbf{VII} we consider the time delay problem in the context of wormholes. Finally in Sec. \textbf{VIII} we comment on our results.

\section{Black hole solution in Massive gravity }

We commence with a brief discussion about black holes in  massive gravity.
An action of a four-dimensional  massive gravity model which is used in this paper,  is given by:
\begin{equation} S = \int  d^4x  \sqrt{-g} \left[ \frac{R}{16 \pi }+ \Lambda^4 \mathcal{F}(X, W^{ij}) \right] ,
\end{equation}
where $R$ is as usual the scalar curvature and $\mathcal{F}$ is a function of the scalar fields $\phi^i$ and $\phi^0$,  which  are minimally coupled to gravity.  These scalar fields play the crucial role  for spontaneously breaking Lorentz
symmetry.  Actually, this  action in  massive gravity  can be treated  as the  low-energy effective theory below the ultraviolet cutoff   $\Lambda$.
The value of $\Lambda$ is of  the order of $\sqrt{mM_{pl}}$ ,  where $m$ is the graviton
mass and $M_{pl}$ is  the Plank mass.
The function $\mathcal{F}$ which depends on two particular combinations of the derivatives of the Goldstone fields, $X$ and $W^{ij}$, are defined as
\begin{equation} X = \frac{\partial^0 \phi^i\partial_0\phi^i}{\Lambda^4},
\end{equation}
\begin{equation}
W^{ij} = \frac{\partial^\mu \phi^i\partial_\mu \phi^i}{\Lambda^4}-\frac{\partial^\mu \phi^i\partial_\mu \phi^0 \partial^\nu \phi^j\partial_\nu \phi^0}{\Lambda^4 X},
\end{equation}
where the constant $\Lambda$ has the dimension of mass. From this, one can arrive at
the new type of black hole solution, namely, massive gravity black hole
 (for detailed derivation can be found in \cite{bebronne}).
The ansatz for the static spherically symmetric black hole solutions can be written in the following form:
 \begin{equation} \label{metric1}
ds^2=- f(r) dt^2+\frac{dr^2}{f(r)}+r^2\left(d\theta^2+\sin^2\theta d\varphi^2   \right),
\end{equation}
where the metric function with the scalar fields are assumed in the following form
  \begin{equation} \label{metric1}
f(r)  =  1-\frac{2M}{r}-\frac{S}{r^{\lambda}},~~~ \phi^0 = \Lambda^2(t+h(r) )~~\text{and}~~\phi^i= \Lambda^2x^i,
\end{equation}
with
 \[h(r) = \pm \int\frac{dr}{f(r)} \left[ 1-f(r) \left( \frac{S\lambda (\lambda-1)}{12m^2} \frac{1}{r^{\lambda+2}} +1\right)^{-1}\right]^{\frac{1}{2}},\]
where $M$ accounts for the gravitational mass of the body and
$\lambda$ is a parameter of the model which depends on the scalar charge $S$. The presence of the
scalar charge represents a modification of the Einstein's gravitational theory.
When $S = 0$ the usual Schwarzschild potential is regained.
However, at large distances with positive $M$ the solution (2) has an attractive behavior,
whereas with negative $M$ the Newton potential is repulsive at large distances and attractive near the horizon.
Our goal is to study the when $M > 0$ and  $S > 0$, so that black hole has attractive gravitational potential at all distances and the size of the event horizon is larger than $2M$. Another reason for considering
such a solution is that the asymptotic behaviour of the gravitational potential is Newtonian
with finite total energy, featuring an asymptotic behavior slower than $1/r$ and generically of the form 1$/r^{\lambda}$.
Therefore, the attraction the modified black hole solution exhibits is stronger than that of the usual Schwarzschild black hole due to the presence of ``hair $\lambda$".

\section{Geodesic equations}

Let us turn our attention to the problem of the deflection angle in massive gravity theory in the framework of the geodesic equations. Recently a new black hole solution in the context of the massive gravity theory was found to be \cite{bebronne}
\begin{equation} \label{metric1}
ds^2=-\left(1-\frac{2M}{r}-\frac{S}{r^{\lambda}}\right)dt^2+\left(1-\frac{2M}{r}-\frac{S}{r^{\lambda}}\right)^{-1}dr^2+r^2\left(d\theta^2+\sin^2\theta d\varphi^2   \right).
\end{equation}

This solution does not describe asymptotically flat space in the case $\lambda <0$. For $\lambda = - 2$ the metric coincides with the familiar Schwarzschild de-Sitter spacetime consisting of a constant stress energy tensor in the form of the (positive) cosmological constant \cite{nariai}.  In the present paper we shall focus on the case $\lambda \geq 1$. Immediately it may be recognized that the case $\lambda = 2$ corresponds with the Reissner--Nordstr\"{o}m solution for the exterior of a charged perfect fluid sphere.   Applying the variational principle to the metric \eqref{metric1} we find the Lagrangian
\begin{eqnarray}\label{lag1}
2\,\mathcal{L}=\left(1-\frac{2M}{r(s)}-\frac{S}{r^{\lambda}(s)}\right)\dot{t}^2(s)+\frac{\dot{r}^2(s)}{\left(1-\frac{2M}{r(s)}-\frac{S}{r^{\lambda}(s)}\right)}+r^2(s)\left(\dot{\theta}^2(s)+\sin^2\theta \dot{\varphi}^2   (s)\right).
\end{eqnarray}

It is worth noting that $\mathcal{L}$ is $+1, 0,$ and $-1$, for timelike, null, and spacelike geodesics, respectively. Taking the equatorial plane $\theta=\pi/2$, the spacetime symmetries implies two constants of motion, namely $l$ and $\mathcal{E}$, given as follows
\begin{eqnarray}
p_{\varphi}&=& \frac{\partial \mathcal{L}}{\partial \dot{\varphi}} = r(s)^2 \dot{\varphi}=l,\\
p_{t}&=& -\frac{\partial \mathcal{L}}{\partial \dot{t}}= \left(1-\frac{2M}{r(s)}-\frac{S}{r^{\lambda}(s)}\right) \dot{t}=\mathcal{E}.
\end{eqnarray}

To proceed further we need to introduce a new variable, say $u(\varphi)$, which is can be given in terms of the radial coordinate as $r=1/u(\varphi)$ which yields the identity
\begin{equation}\label{iden1}
\frac{\dot{r}}{\dot{\varphi}}=\frac{\mathrm{d}r}{\mathrm{d}\varphi}=-\frac{1}{u^2}\frac{\mathrm{d}u}{\mathrm{d}\varphi}
\end{equation}

After some algebraic manipulations one can show that the following relation  can be recovered
\begin{eqnarray}\label{diff2}
-\frac{\dot{t}^2(s)}{\dot{\varphi}^2(s)}+2 \frac{\dot{t}^2(s)}{\dot{\varphi}^2(s)} Mu+Su^{\lambda}\frac{\dot{t}^2(s)}{\dot{\varphi}^2(s)}+\left(\frac{du}{d\varphi}\right)^2 \frac{1}{u^4 \left(1-2Mu-Su^{\lambda{}}  \right)}+\frac{1}{u^2}=0.
\end{eqnarray}

On the other hand, from Eqs. (8) and (9) we find
\begin{eqnarray}
\frac{\dot{t}(s)}{\dot{\varphi}(s)}=\frac{\mathcal{E}}{l\left(1-2Mu-\frac{S}{u^{\lambda}}  \right)u^2}.
\end{eqnarray}

Hence we can recast Eq. (11) in terms of the impact parameter $b$ as follows
\begin{eqnarray}\label{diff2}
\frac{\left(2Mu+Su^\lambda-1 \right)}{b^2\left(1-2Mu-\frac{S}{u^{\lambda}}  \right)^2u^4}+\left(\frac{du}{d\varphi}\right)^2 \frac{1}{u^4 \left(1-2Mu-Su^{\lambda{}}  \right)}+\frac{1}{u^2}=0.
\end{eqnarray}
where $b$ is defined as
\begin{equation}
b=\frac{l}{\mathcal{E}}.
\end{equation}

We proceed by considering four special cases for different values of the parameter $\lambda$ in the metric (6).

\subsection{Case $\lambda=1$}
To begin, we shall consider the affine parameter along the light rays to be  $\mathcal{E}=1$, therefore one should find the following condition $ u_{max} = 1/r_0$, where $r_0$ gives the distance of the closest approach. Next, we can evaluate the constant $l$ from Eq. (14) in leading order terms as
\begin{equation}
l=\left(\sqrt{\frac{4MS}{r_{0}^2}+\frac{S}{r_{0}}+\frac{2M}{r_{0}}+1} \right)\,r_{0}.
\end{equation}

This leads us to the following differential equation
\begin{equation}
\left(\frac{du}{d\varphi}\right)^2\frac{1}{u^4 \mathcal{K}}+\frac{1}{u^4 r_0^2 \Upsilon^2 \mathcal{K}^2}-\frac{2M}{u^3 r_0^2 \Upsilon^2 \mathcal{K}^2}-\frac{S}{r_0^2 \Upsilon^2 \mathcal{K}^2}-\frac{1}{u^2}=0,
\end{equation}
where
\begin{eqnarray}
\mathcal{K}&=& Su+2Mu-1, \\
\Upsilon &=& \frac{4MS}{r_0^2}+\frac{S}{r_0}+\frac{2M}{r_0}+1.
\end{eqnarray}

From the above equation we find
\begin{equation}
\frac{d\varphi}{du}=\pm \sqrt{\frac{\mathcal{C}_{1}}{\mathcal{A}_{1}u^3-\mathcal{C}_{1}u^2+1}},
\end{equation}
where
\begin{eqnarray}
\mathcal{A}_{1}&=& 2M^2 S+4M^2 r_0+4 M S^2+4MSr_0+2Mr_0^2+S^2r_0+Sr_0^2,\\
\mathcal{C}_{1}&=& 2Mr_0^2+4MS+Sr_0+r_0^2.
\end{eqnarray}

It is well known that the solution to the above equation in the weak limit can be written as follows \cite{weinberg1} 
\begin{equation}
\Delta \varphi =\pi+\hat{\alpha},
\end{equation}
where $\hat{\alpha}$ is the deflection angle which should be calculated. Moreover, from the above equation the deflection angle is shown to be calculated  as follows \cite{weinberg1}
\begin{equation}
\hat{\alpha}=2|\varphi_{u={1/b}}-\varphi_{u=0}|-\pi.
\end{equation}

Using this relation, from Eq. (19) the deflection angle is found to be
\begin{equation}
\hat{\alpha}_{\lambda=1} \simeq \frac{4M}{r_0}+\frac{M^2}{r_0^2}\left(\frac{15 \pi}{4}-4\right)+\frac{2\, S}{r_0}-\frac{MS}{r_0^2}\left(4-\frac{15 \pi}{4} \right)-\frac{S^2}{r_0^2}\left(1-\frac{15 \pi}{16}\right).
\end{equation}

Furthermore if we let $S=0$, we find the Schwarzschild deflection angle with second-order correction terms which is in perfect agreement with \cite{will}.

\subsection{Case $\lambda=2$}
Our second case will be $\lambda=2$. Going through the same procedure as in the last example the constant $l$ is found to be
\begin{equation}
l=\left(\sqrt{\frac{4MS}{r_0^3}+\frac{S}{r_0^2}+\frac{2M}{r_0}+1} \right)\,r_0.
\end{equation}

We obtain the following differential equation
\begin{equation}
\left(\frac{du}{d\varphi}\right)^2\frac{1}{u^4 \mathcal{M}}+\frac{1}{u^4 r_0^2 \Delta^2 \mathcal{M}^2}-\frac{2M}{u^3 r_0^2 \Delta^2 \mathcal{M}^2}-\frac{S}{r_0^2 \Delta^2 \mathcal{M}^2}-\frac{1}{u^2}=0,
\end{equation}
where
\begin{eqnarray}
\mathcal{M}&=& Su^2+2Mu-1, \\
\Delta &=& \frac{4MS}{r_0^3}+\frac{S}{r_0^2}+\frac{2M}{r_0}+1.
\end{eqnarray}

From the above equation we get that
\begin{equation}
\frac{d\varphi}{du}=\pm \sqrt{\frac{\mathcal{C}_{2}}{\mathcal{A}_{2}u^4+\mathcal{B}_{2}u^3-\mathcal{C}_{2}u^2+r_0}},
\end{equation}
where
\begin{eqnarray}
\mathcal{A}_{2}&=& 2M Sr_0^2+Sr_0^3+4MS^2+S^2r_0,\\
\mathcal{B}_{2}&=& 4M^2 r_0^2+2Mr_0^3+8 M^2 S+2MS r_0,\\
\mathcal{C}_{2}&=& 2Mr_0^2+4MS+Sr_0+r_0^3.
\end{eqnarray}

Consequently the deflection angle has the form
\begin{equation}
\hat{\alpha}_{\lambda=2} \simeq \frac{4M}{r_0}+\frac{M^2}{r_0^2}\left(\frac{15 \pi}{4}-4\right)+\frac{3\, S\,\pi}{4 \,r_0^2}+\frac{MS}{r_0^3}\left(14-\frac{3 \pi}{2} \right)+\frac{57\,\pi\, S^2}{64\, r_0^4}
\end{equation}

Now as a special case we can find the charged black hole deflection angle by simply letting $S=-Q^2$. In that case we find the RN deflection angle
\begin{equation}
\hat{\alpha}_{RN} \simeq \frac{4M}{r_{0}}+\frac{M^2}{r_{0}^2}\left(\frac{15 \pi}{4}-4\right)-\frac{3\, Q^2\,\pi}{4 \,r_{0}^2}-\frac{MQ^2}{r_{0}^3}\left(14-\frac{3 \pi}{2} \right)+\frac{57\,\pi\, Q^4}{64\, r_{0}^4}
\end{equation}

\subsection{Case $\lambda=3$}
In a similar way, letting $\lambda=3$ we found
\begin{equation}
l=\left(\sqrt{\frac{4MS}{r_0^4}+\frac{S}{r_0^3}+\frac{2M}{r_0}+1} \right)\,r_0.
\end{equation}

The differential equation takes the form
\begin{equation}
\left(\frac{du}{d\varphi}\right)^2\frac{1}{u^4 \mathcal{N}}+\frac{1}{u^4 r_0^2 \Theta \mathcal{N}^2}-\frac{2M}{u^3 r_0^2 \Theta \mathcal{N}^2}-\frac{S}{r_0^2 \Theta \mathcal{N}^2}-\frac{1}{u^2}=0,
\end{equation}
where
\begin{eqnarray}
\mathcal{N}&=& Su^3+2Mu-1, \\
\Theta &=& \frac{4MS}{r_0^4}+\frac{S}{r_0^3}+\frac{2M}{r_0}+1.
\end{eqnarray}

From the above equation we find
\begin{equation}
\frac{d\varphi}{du}=\pm \sqrt{\frac{\mathcal{C}_{3}}{\mathcal{A}_{3}u^5+\mathcal{B}_{3}u^3-\mathcal{C}_{3}u^2+r_0^2}},
\end{equation}
where
\begin{eqnarray}
\mathcal{A}_{3}&=& 2M Sr_0^3+Sr_0^4+4MS^2+S^2r_0,\\
\mathcal{B}_{3}&=& 4M^2 r_0^3+2Mr_0^4+8 M^2 S+2MS r_0,\\
\mathcal{C}_{3}&=& 2Mr_0^3+4MS+Sr_0+r_0^4.
\end{eqnarray}

The deflection angle is given by
\begin{equation}
\hat{\alpha}_{\lambda=3} \simeq \frac{4M}{r_0}+\frac{M^2}{r_0^2}\left(\frac{15 \pi}{4}-4\right)+\frac{8\, S}{3 \,r_0^3}+\frac{MS}{r_0^4}\left(10+\frac{105 \pi}{16} \right)+\frac{315 \,\pi\, S^2}{128\, r_0^6}
\end{equation}

\subsection{Case $\lambda=4$}
Finally, in our last case we let  $\lambda=4$, it follows
\begin{equation}
l=\left(\sqrt{\frac{4MS}{r_0^5}+\frac{S}{r_0^4}+\frac{2M}{r_0}+1} \right)\,r_0.
\end{equation}

We find the following differential equation
\begin{equation}
\left(\frac{du}{d\varphi}\right)^2\frac{1}{u^4 \Xi}+\frac{1}{u^4 r_0^2 \zeta \Xi^2}-\frac{2M}{u^3 r_0^2 \zeta \Xi^2}-\frac{S}{r_0^2 \zeta \Xi^2}-\frac{1}{u^2}=0,
\end{equation}
where
\begin{eqnarray}
\Xi&=& Su^4-2Mu-1, \\
\zeta &=& \frac{4MS}{r_0^5}+\frac{S}{r_0^4}+\frac{2M}{r_0}+1.
\end{eqnarray}

From the above equation we obtain
\begin{equation}
\frac{d\varphi}{du}=\pm \sqrt{\frac{\mathcal{C}_{4}}{\mathcal{A}_{4}u^6+\mathcal{B}_{4}u^3-\mathcal{C}_{4}u^2+r_0^3}},
\end{equation}
where
\begin{eqnarray}
\mathcal{A}_{4}&=& 2M Sr_0^4+Sr_0^5+4MS^2+S^2r_0,\\
\mathcal{B}_{4}&=& 4M^2 r_0^4+2Mr_0^5+8 M^2 S+2MS r_0,\\
\mathcal{C}_{4}&=& 2Mr_0^4+4MS+Sr_0+r_0^5 .
\end{eqnarray}

Expanding in Taylor series and integrating we derive the expression
\begin{equation}
\hat{\alpha}_{\lambda=4} \simeq \frac{4M}{r_0}+\frac{M^2}{r_0^2}\left(\frac{15 \pi}{4}-4\right)+\frac{15 \pi S}{16 \,r_0^4}+\frac{MS}{r_0^5}\left( \frac{118}{5}-\frac{15 \pi}{4} \right)+\frac{1545 \,\pi\, S^2}{1024\, r_0^8}
\end{equation}

\begin{figure}[h!]
\center
\includegraphics[width=0.45\textwidth]{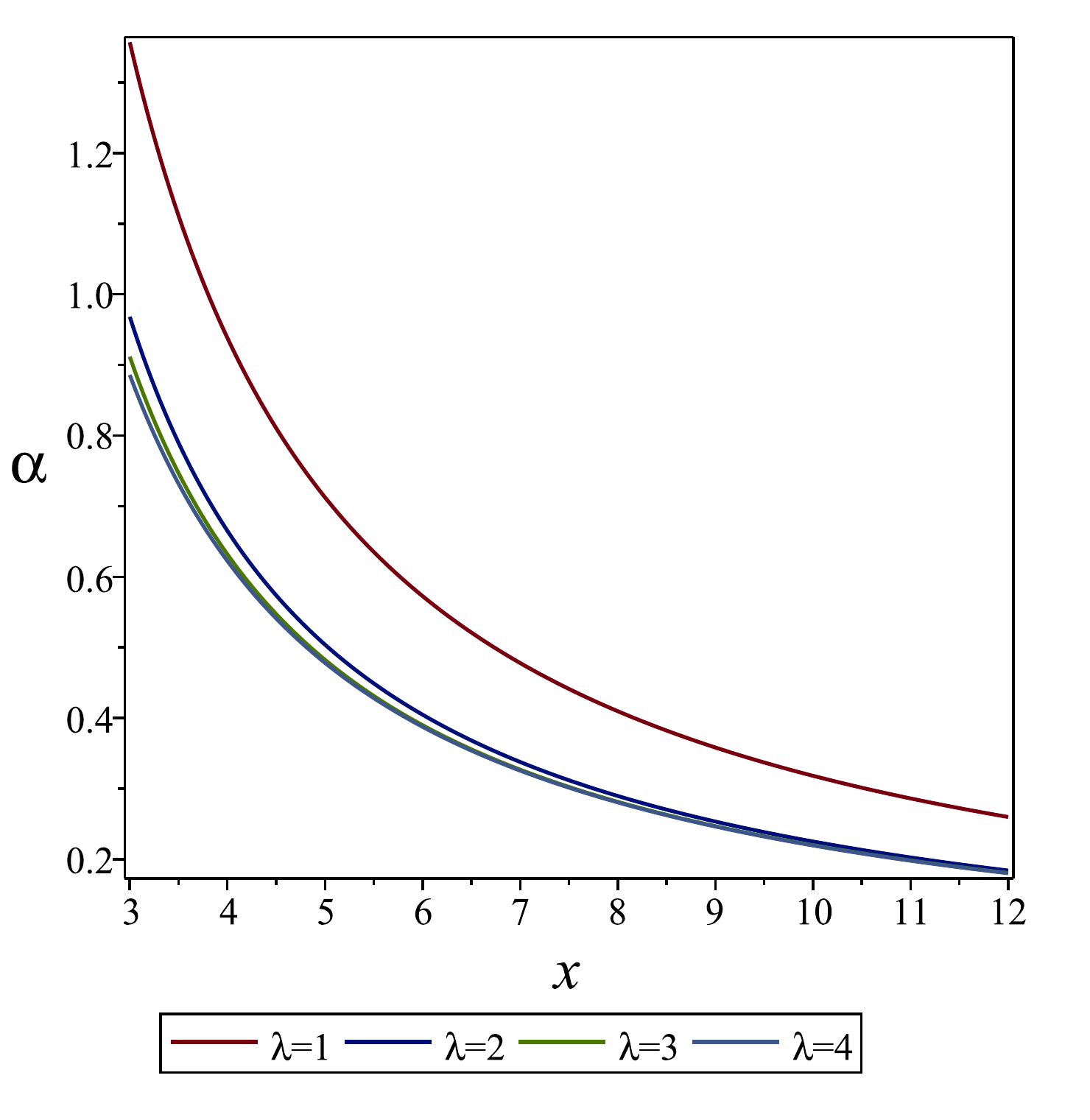}
\includegraphics[width=0.45\textwidth]{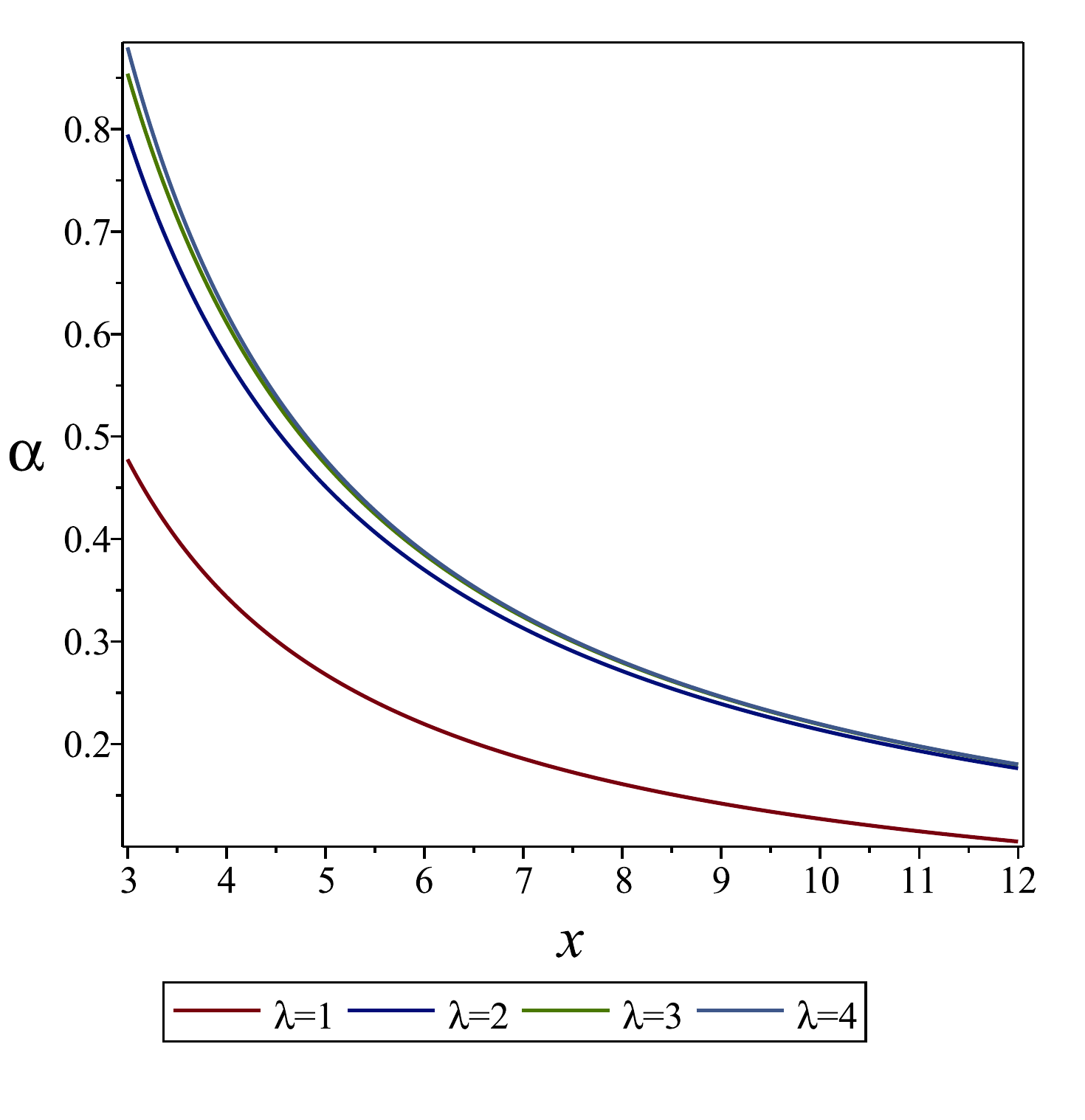}
\caption{\small \textit{We plot the deflection angle as a function of $x=r_0/2M$. In the first plot we have chosen $M=1$ and $S=0.8$. One can observe that with the increase of $\lambda$ the deflection angle decreases for fixed valued of $M$ and the scalar charge being positive i.e. $S>0$. In the second plot we choose $M=1$ and $S=-0.8$. In this case with the increase of $\lambda$ the deflection angle increases. }}
\end{figure}

\section{Gauss-Bonnet (GB) method}
\subsection{Gaussian optical curvature}

In this subsection we consider null geodesics deflected by a
black hole in massive gravity models. We start by considering the optical metric from spacetime metric (6), by choosing the null geodesic equations $ds^2$ = 0.  In the equatorial plane $\theta = \pi/2$ we find
\begin{equation}\label{dt}
dt^2= \frac{dr^2}{(1-\frac{2M}{r}-\frac{S}{r^{\lambda}})^{2}}+\frac{r^{2}}
{(1-\frac{2M}{r}-\frac{S}{r^{\lambda}})}d\varphi^{2}\equiv dr^{\star{2}}+(f(r^{\star}))^{2}d\varphi^{2}.
\end{equation}

For the following considerations, it is convenient to introduce a radial Regge-Wheeler tortoise coordinate $r^\star$, with a new function  $f(r^\star)$ as follows:
\begin{eqnarray}
dr^\star &=&\frac{dr}{(1-\frac{2M}{r}-\frac{S}{r^{\lambda}})},
\end{eqnarray}
\begin{eqnarray}
f(r^\star)&=& \frac{r}{(1-\frac{2M}{r}-\frac{S}{r^{\lambda}})^{\frac{1}{2}}}.
\end{eqnarray}

This prescription allows us to write the  line element of the optical metric in the form
\begin{equation}
dt^2 \equiv g_{ab}^{op} dx^a dx^b={dr^{\star}}^{2}+{f(r^{\star})}^2 d\varphi^2.
\end{equation}

Using this static coordinates system, it is now clear that the equatorial plane in the
optical metric is a surface of revolution when it is embedded, in $\mathbb{R}^3$.
We utilized the following mathematical formulae to calculate the Gaussian curvature
$K$, of the optical surface as
\begin{eqnarray}
K&=&-\frac{1}{f(r^{\star })}\frac{d^{2}f(r^{\star })}{d{%
r^{\star }}^{2}}\\
&=& -\frac{1}{f(r^{\star })}\left[ \frac{dr}{dr^{\star }}%
\frac{d}{dr}\left( \frac{dr}{dr^{\star }}%
\right) \frac{df}{dr}+\left( \frac{dr}{d%
r^{\star }}\right) ^{2}\frac{d^{2}f}{dr^{2}}\right].\notag
\end{eqnarray}
With the help of  Eq. (54) the optical Gaussian curvature may be expressed as
 (for further review see \cite{gibbons1})
\begin{eqnarray}
K&=&\frac{\lambda(\lambda+2)S^{2}}{4r^{2\lambda+2}}+\frac{(\lambda^{2}+2)SM}{r^{\lambda+3}}
-\frac{\lambda(\lambda+1)S}{2r^{\lambda+2}}-M(\frac{2}{r^{3}}-\frac{3M}{r^{4}}).
\end{eqnarray}


\subsection{Deflection angle}
\textbf{Theorem:} \textit{Let  $\mathcal{S}_{R}$ be a non-singular region with boundary $\partial
\mathcal{S}_{R}=\gamma _{g^{op}}\cup \gamma_{R}$, and let $K$ and $\kappa$ be the Gaussian optical curvature and the geodesic curvature, respectively. Then GBT reads}  \cite{gibbons1}
\begin{equation}\label{GBT}
\iint\limits_{\mathcal{S}_{R}}K\,dA+\oint\limits_{\partial \mathcal{%
S}_{R}}\kappa \,dt+\sum_{i}\theta _{i}=2\pi \chi (\mathcal{S}_{R}),
\end{equation}
in  which $\theta_{i}$  are the exterior angles at the $i^{th}$ vertex. In our setup, however,  the Euler characteristic is  $\chi (\mathcal{S}_{R}) = 1$ due to the fact that we consider a non-singular domain outside of the light ray. It is worth noting that for a singular domain  we have $\chi (\mathcal{S}_{R}) = 0$.

Furthermore, for computing the deflection angle of light,  we need first to compute the geodesic curvature in terms of the following relation
\begin{equation}
\kappa =g^{\text{op}}\,\left(\nabla _{\dot{%
\gamma}}\dot{\gamma},\ddot{\gamma}\right).
\end{equation}

In doing so we should take into account the unit speed condition which is stated as follows $g^{op}(\dot{\gamma},\dot{%
\gamma}) = 1$, with $\ddot{\gamma}$ being the unit acceleration vector. Next, if we simply allow $R\rightarrow \infty $, one can show that our two jump angles ($\theta _{\mathcal{O}}$, $\theta _{\mathcal{S}}$) yield $\pi /2$. Put it differently, if we take the total sum of our jump angles at $\mathcal{S}$ and $\mathcal{O}$, we find $\theta _{\mathit{O}}$ + $ \theta _{\mathit{S}}\rightarrow \pi$ \cite{gibbons1}. It follows from the simple geometry that $\kappa (\gamma _{g^{op}})=0$ due to the simple fact that $\gamma _{g^{op}}$ is a geodesic. Hence we are left with the following relation
\begin{equation}
\kappa (\gamma_{R})=|\nabla _{\dot{\gamma}_{R}}\dot{\gamma}_{R}|,
\end{equation}
in which  $\gamma_{R}:=r(\varphi)=R=\text{constant}$. In this way, one is left with the following non-zero radial part 
\begin{equation}
\left( \nabla _{\dot{\gamma}_{R}}\dot{\gamma}_{R}\right) ^{r}=\dot{\gamma}_{R}^{\varphi
}\,\left( \partial _{\varphi }\dot{\gamma}_{R}^{r}\right) +\tilde{\Gamma} _{\varphi
\varphi }^{r}\left( \dot{\gamma}_{R}^{\varphi }\right) ^{2}, \label{12}
\end{equation}
note that $\tilde{\Gamma}_{\varphi\varphi }^{r}$ is the Christoffel symbol associated with the optical metric geometry. While is clear that the first term in this equation must vanish, we can calculate the second term via the condition $\tilde{g}_{\varphi \varphi}\dot{\gamma}_{R}^{\varphi } \dot{\gamma}_{R}^{\varphi }=1 $. Finally we find 
\begin{eqnarray}\notag\label{gcurvature}
\lim_{R\rightarrow \infty }\kappa (\gamma_{R}) &=&\lim_{R\rightarrow \infty
}\left\vert \nabla _{\dot{\gamma}_{R}}\dot{\gamma}_{R}\right\vert , \notag \\
&\rightarrow &\frac{1}{R}.
\end{eqnarray}%

But for very large radial distance Eq. \eqref{dt}, suggest that
\begin{eqnarray}\notag\label{dt1}
\lim_{R\rightarrow \infty } dt &=&\lim_{R\rightarrow \infty
}\left[\frac{R }{\left(1-\frac{2M}{R}-\frac{S}{R^{\lambda}}\right)^{1/2}}\right] d \varphi\\
&\to &  R \, d \varphi,
\end{eqnarray}%
provided that $\lambda >0$. From GBT we find
\begin{equation}
\iint\limits_{\mathcal{S}_{R}}K\,dA+\oint\limits_{\gamma_{R}}\kappa \,%
\mathrm{d}t\overset{{R\rightarrow \infty }}{=}\iint\limits_{\mathcal{S}%
_{\infty }}K\,dA+\int\limits_{0}^{\pi + \hat{\alpha}}d\varphi
=\pi,
\end{equation}
where the surface element is given by $dA=\sqrt{\det g^{op}}\,dr^{\star}d\varphi.$ It is clear now that we should integrate over the domain $\mathcal{S}_{\infty}$ to find the deflection angle. This the deflection angle is found to be
\begin{eqnarray}\label{GBT2}
\hat{\alpha}^{GB}&=&-\int\limits_{0}^{\pi}\int\limits_{r_{\gamma}}^{\infty} K \sqrt{\det g^{op}}\,dr^{\star}d\varphi.
\end{eqnarray}

One can now compute the deflection angle by choosing the light ray as $ r(\varphi) =b/ \sin\varphi$. However, this equation corresponds to the straight-line approximation and gives the correct result only for the linear terms in the deflection angle. In this paper, we will make use of the following choice for the light ray which is a solution of our geodesic equation (13):
\begin{equation}
\frac{1}{r_{\gamma}}={\frac {\sin \left( \varphi \right) }{b}}+\frac{1}{2}{\frac {M \left( 3+\cos
 \left( 2\,\varphi \right)  \right) }{{b}^{2}}}+\frac{1}{16}{\frac {{M}^{2}
 \left( 37\,\sin \left( \varphi \right) +30\, \left( \pi -2\,\varphi
 \right) \cos \left( \varphi \right) -3\,\sin \left( 3\,\varphi \right)
 \right) }{b}}.
\end{equation}

Let us now elaborate on the following special cases:

\subsubsection{$\lambda=1$}
Let us first calculate the Gaussian optical curvature from Eq. (58) in the case when $\lambda=1$. One can easily find that
\begin{equation}
K_{\lambda=1} \simeq -\frac {2\,M+S}{{r}^{3}}+\frac { \left( 3\,S+6\,M \right)
 \left( 2\,M+S \right) }{4\,{r}^{4}}.
\end{equation}

Substituting into Eq. \eqref{GBT2} generates the value of the deflection angle in terms of the integral
\begin{eqnarray}
\hat{\alpha}_{\lambda=1}^{GB}&=&-\int\limits_{0}^{\pi}\int\limits_{r_{\gamma}}^{\infty}\left( -{\frac {2\,M+S}{{r}^{3}}}+{\frac { \left( 3\,S+6\,M \right)
 \left( 2\,M+S \right) }{4\,{r}^{4}}}
\right)\sqrt{\det g^{op}}\,dr^{\star}d\varphi.
\end{eqnarray}

In order to evaluate the above integral  note that
\begin{equation}
\sqrt{\det g^{op}}\,dr^{\star}=\frac{r\, dr}{\left(1-\frac{2M}{r}-\frac{S}{r}  \right)^{3/2}}
\end{equation}
and expanding  in  a Taylor series the previous equation results in the expression
\begin{equation}
\sqrt{\det g^{op}}\,dr^{\star} = r\, dr\left[1+\frac{3M}{r}+\frac{15 M^2}{2r^2} +\left(\frac{3}{2r}+\frac{15M}{2r^2}\right)S+...   \right]
\end{equation}

Using the above result for the deflection angle we find
\begin{equation}
\hat{\alpha}_{\lambda=1}^{GB} \simeq \frac{4M}{b}+\frac{15 M^2\pi}{4\,b^2}+\frac{2S}{b}+\frac{9 \pi MS}{4 b^2}+\frac{3 \pi S^2}{16 b^2}.
\end{equation}

On the other hand we can use the relation (15) to express the last result in terms of the minimal distance $r_0$ in terms of the impact parameter
\begin{equation}
\frac{1}{b}= \frac{1}{r_0}\left(1-\frac{M}{r_0}-\frac{S}{2 r_0}+...\right)
\end{equation}

Consequently the deflection angle takes the form
\begin{equation}
\hat{\alpha}_{\lambda=1}^{GB} \simeq \frac{4M}{r_0}+\frac{M^2}{r_0^2}\left(\frac{15 \pi}{4}-4\right)+\frac{2S}{r_0}-\frac{S^2}{r_0^2}\left(1- \frac{3 \pi}{16}\right).
\end{equation}

Thus we have shown  that by modifying the integration domain our result is in perfect agreement up to the second order in $M$, and agrees only in the linear term in $S$. In order to find the exact result including the second order terms in $S$ we have to modify the equation for the light ray (65). However this goes beyond the scope of this paper.

\subsubsection{$\lambda=2$}
Let us substitute this equation into Eq. \eqref{GBT2} then we find that the deflection angle is given in terms of the following integral
\begin{eqnarray}
\hat{\alpha}_{\lambda=2}&=&-\int\limits_{0}^{\pi}\int\limits_{r_{\gamma}}^{\infty}\left( -2\,{\frac {M}{{r}^{3}}}+{\frac {3\,{M}^{2}-3\,S}{{r}^{4}}}+6\,{\frac
{MS}{{r}^{5}}}+2\,{\frac {{S}^{2}}{{r}^{6}}}
\right)\sqrt{\det g^{op}}\,dr^{\star}d\varphi.
\end{eqnarray}
where
\begin{equation}
\sqrt{\det g^{op}}\,dr^{\star} = r\, dr\left[1+\frac{3M}{r}+\frac{15 M^2}{2r^2} +\left(\frac{3}{2r^2}+\frac{15M}{2r^3}\right)S+...   \right]
\end{equation}

The deflection angle in terms of the impact factor is found to be
\begin{equation}
\hat{\alpha}_{\lambda=2}^{GB} \simeq \frac{4M}{b}+\frac{15 M^2\pi}{4\,b^2}+\frac{3 \pi S}{4 \, b^2}+\frac{32 MS}{3 \,b^3}+\frac{15 \pi S^2}{64\, b^4}.
\end{equation}

As already noted, the disagreement in the last two terms is to be expected due to the integration domain. Finally, neglecting these terms and letting $S=-Q^2$, if we expand (25) in series form the last result we recover Eq. (34) up to the second order terms in $M$ and $Q$.

\subsubsection{$\lambda=3$}
Let us substitute this equation into Eq. \eqref{GBT2} then we find that the deflection angle is given in terms of the following integral
\begin{eqnarray}
\hat{\alpha}_{\lambda=3}&=&-\int\limits_{0}^{\pi}\int\limits_{r_{\gamma}}^{\infty}\left(3\,{\frac {{M}^{2}}{{r}^{4}}}-2\,{\frac {M}{{r}^{3}}}+11\,{\frac {MS}{
{r}^{6}}}-6\,{\frac {S}{{r}^{5}}}+{\frac {15\,{S}^{2}}{4\,{r}^{8}}}
\right)\sqrt{\det g^{op}}\,dr^{\star}d\varphi.
\end{eqnarray}
where
\begin{equation}
\sqrt{\det g^{op}}\,dr^{\star} = r\, dr\left[1+\frac{3M}{r}+\frac{15 M^2}{2r^2} +\left(\frac{3}{2r^3}+\frac{15M}{2r^4}\right)S+...   \right]
\end{equation}

The deflection angle has the form
\begin{equation}
\hat{\alpha}_{\lambda=3}^{GB} \simeq \frac{4M}{b}+\frac{15 M^2\pi}{4\,b^2}+\frac{8 S}{3 b^3}+\frac{75 \pi MS }{16\, b^4}+\frac{35 \pi S^2}{128 \,b^6}.
\end{equation}

Hence in a similar way using Eq. (35) we recover Eq. (43) up to the second order in $M$, but in leading order in $S$.

\subsubsection{$\lambda=4$}

Let us substitute this equation into Eq. \eqref{GBT2} then we find that the deflection angle is given in terms of the following integral
\begin{eqnarray}
\hat{\alpha}_{\lambda=4}&=&-\int\limits_{0}^{\pi}\int\limits_{r_{\gamma}}^{\infty}\left( -2\,{\frac {M}{{r}^{3}}}+3\,{\frac {{M}^{2}}{{r}^{4}}}-10\,{\frac {S}{
{r}^{6}}}+18\,{\frac {MS}{{r}^{7}}}+6\,{\frac {{S}^{2}}{{r}^{10}}}
\right)\sqrt{\det g^{op}}\,dr^{\star}d\varphi.
\end{eqnarray}
where
\begin{equation}
\sqrt{\det g^{op}}\,dr^{\star} = r\, dr\left[1+\frac{3M}{r}+\frac{15 M^2}{2r^2} +\left(\frac{3}{2r^4}+\frac{15M}{2r^5}\right)S+...   \right]
\end{equation}

The deflection angle is given by
\begin{equation}
\hat{\alpha}_{\lambda=4}^{GB} \simeq \frac{4M}{b}+\frac{15 M^2\pi}{4\,b^2}+\frac{15 \pi S}{16\, b^4}+\frac{96 MS}{5 b^5}+\frac{315 \pi S^2}{1024}.
\end{equation}

Or, after we use Eq. (44) the deflection angle in terms of the distance of the closest approach reads
\begin{equation}
\hat{\alpha}_{\lambda=4}^{GB} \simeq \frac{4M}{r_0}+\frac{M^2}{r_0^2}\left(\frac{15 \pi}{4}-4\right)+\frac{15 \pi S}{16 \,r_0^4}.
\end{equation}

\section{Time Delay}
We analyze here the  time delay due to the massive gravitational field of the black hole solution. Suppose that two photons emitted at the same time but follow different paths to reach the observer. They will take two different times to reach the observer and this time difference is called the \textit{time delay}. It is important to discuss the time delays between lensed multiple images which is directly related to determining the Hubble constant $H_{0}$  and was first pointed out by Refsdal \cite{Refsdal}.

We consider light propagation in a static spherically symmetric spacetime  given by the line element
\begin{equation}
ds^2 = -A(r) dt^2 + B(r) dr^2 + C(r)(d\theta^2+\sin^2\theta d\phi^2).
\end{equation}
The time delay of a light signal passing through the gravitational field of this configuration is express as
\begin{eqnarray}
\Delta T = 2\int_{r_{0}}^{r_{1}}\left[\frac{1}{\sqrt{\left[\frac{A(r)}{B(r)}-\frac{A^2(r)}{B(r)C(r)}\frac{C(r_0)}{A(r_0)}\right]}}-\frac{1}{\sqrt{\left[1-\frac{r_0^2}{r^2}\right]}}\right]dr\nonumber\\
+ 2\int_{r_{0}}^{r_{2}}\left[\frac{1}{\sqrt{\left[\frac{A(r)}{B(r)}-\frac{A^2(r)}{B(r)C(r)}\frac{C(r_0)}{A(r_0)}\right]}}-\frac{1}{\sqrt{\left[1-\frac{r_0^2}{r^2}\right]}}\right]dr,
\end{eqnarray}
where $r_1$ and $r_2$ are distances of the observer and the source from the configuration and $r_0$ is the closest approach to the configuration. With help of this algorithm we will calculate the time delay due to the massive gravitational field of the black hole.\\
Let $r _{e} $ and $r _{s} $ be distances of the observer (Earth) and the source from the black hole respectively. Further $r _{0} $ is the closest approach to the black hole.

Therefore, the total time required for a light signal passing through the gravitational field of the black hole to go from the observer (Earth) to the source and back after reflection from  the source is given by the following equation \cite{weinberg1}.  \\
\begin{equation}
T _{e},_{s}  = 2\left[t(r _{e} ,r _{0} ) + t(r _{s} ,r _{0} )\right],
\end{equation}
where
\begin{equation}
t(r_{e},r_{0})=\int_{r_{0}}^{r_{e}}\left(1-\frac{2M}{r}-\frac{S}{r^{\lambda}}\right)^{-1}\left(1-\frac{\left(1-\frac{M}{r}-\frac{S}{r^{\lambda}}\right)}{\left(1-\frac{M}{r_{0}}-\frac{S}{r_{0}^{\lambda}}\right)}\frac{r _{0} ^{2}}{r^{2}}\right)^{-\frac{1}{2}}dr,
\end{equation}
and
\begin{equation}
t(r_{s},r_{0})=\int_{r_{0}}^{r_{s}}\left(1-\frac{2M}{r}-\frac{S}{r^{\lambda}}\right)^{-1}\left(1-\frac{\left(1-\frac{M}{r}-\frac{S}{r^{\lambda}}\right)}{\left(1-\frac{M}{r_{0}}-\frac{S}{r_{0}^{\lambda}}\right)}\frac{r _{0} ^{2}}{r^{2}}\right)^{-\frac{1}{2}}dr,
\end{equation}
for our considered metric, given in the Eq. (6).

Considering the  approximations (as $r _{e} $,$r _{s} $, $r _{0} $ $ > >  2M$) the integrand of these expressions  \\

\begin{equation}
\alpha =\left (1-\frac{2M}{r}-\frac{S}{r^{\lambda}}\right) ^{-1}\left(1-\frac{\left(1-\frac{M}{r}-\frac{S}{r^{\lambda}}\right)}{\left(1-\frac{M}{r_{0}}-\frac{S}{r_{0}^{\lambda}}\right)}\frac{r _{0} ^{2}}{r^{2}}\right)^{-\frac{1}{2}}
\end{equation}

assume the  form\\
\begin{equation}
\alpha \approx \left(1-\frac{r _{0} ^{2}}{r^{2}}\right)^{-\frac{1}{2}}\left[1+\frac{2M}{r}+\frac{S}{r^{\lambda}}+\frac{Mr _{0} }{r(r+r _{0} )}+\frac{S(1+\frac{r_{0}}{r}+(\frac{r_{0}}{r})^{2}+.......(\frac{r_{0}}{r})^{\lambda-1})}{2r(r+r_{0})r_{0}^{\lambda-2}}\right]
\end{equation}
 So, we can express the Eq. (85) as
\begin{equation*}
T _{e},_{s} = 2\int_{r_{0}}^{r_{e}}\left(1-\frac{r _{0} ^{2}}{r^{2}}\right)^{-\frac{1}{2}}\left[1+\frac{2M}{r}+\frac{S}{r^{\lambda}}+\frac{Mr _{0} }{r(r+r _{0} )}+\frac{S\left(1+\frac{r_{0}}{r}+(\frac{r_{0}}{r})^{2}+.......(\frac{r_{0}}{r})^{\lambda-1}\right)}{2r(r+r_{0})r_{0}^{\lambda-2}}\right]dr
\end{equation*}
 \begin{equation}
+ 2\int_{r_{0}}^{r_{s}}\left(1-\frac{r _{0} ^{2}}{r^{2}}\right)^{-\frac{1}{2}}\left[1+\frac{2M}{r}+\frac{S}{r^{\lambda}}+\frac{Mr _{0} }{r(r+r _{0} )}+\frac{S\left(1+\frac{r_{0}}{r}+(\frac{r_{0}}{r})^{2}+.......(\frac{r_{0}}{r})^{\lambda-1}\right)}{2r(r+r_{0})r_{0}^{\lambda-2}}\right]dr
\end{equation}

 In the absence of  gravitational field ($M = S = 0$) the time is \\
\begin{equation}
T^{'}_{e},_{s}= 2\left[\int_{r_{0}}^{r_{e}}\left(1-\frac{r _{0} ^{2}}{r^{2}}\right)^{-\frac{1}{2}}+\int_{r_{0}}^{r_{s}}\left(1-\frac{r _{0} ^{2}}{r^{2}}\right)^{-\frac{1}{2}}\right].
\end{equation}

 Now, the  delay in time is express as the following equation
\begin{equation}
\Delta T _{e},_{s} = T _{e},_{s} - T^{'}_{e},_{s}.
\end{equation}

Finally, we can estimate  the time delay due to the gravitational field of the  black hole as
\begin{equation*}
\Delta T _{e},_{s}= 2\int_{r_{0}}^{r_{e}}\left(1-\frac{r _{0} ^{2}}{r^{2}}\right)^{-\frac{1}{2}}\left[\frac{2M}{r}+\frac{S}{r^{\lambda}}+\frac{Mr _{0} }{r(r+r _{0} )}+\frac{S\left(1+\frac{r_{0}}{r}+(\frac{r_{0}}{r})^{2}+.......(\frac{r_{0}}{r})^{\lambda-1}\right)}{2r(r+r_{0})r_{0}^{\lambda-2}}\right]dr
\end{equation*}

\begin{equation}
 + 2\int_{r_{0}}^{r_{s}}\left(1-\frac{r _{0} ^{2}}{r^{2}}\right)^{-\frac{1}{2}}\left[\frac{2M}{r}+\frac{S}{r^{\lambda}}+\frac{Mr _{0} }{r(r+r _{0} )}+\frac{S\left(1+\frac{r_{0}}{r}+(\frac{r_{0}}{r})^{2}+.......(\frac{r_{0}}{r})^{\lambda-1}\right)}{2r(r+r_{0})r_{0}^{\lambda-2}}\right]dr,
\end{equation}
and we may proceed to calculate the delay in time for the  cases corresponding to the values of $\lambda = 1, 2, 3$,  and 4 respectively.

\subsection{Case $\lambda$=1}

\begin{eqnarray}
\Delta T _{e},_{s}\mid_{\lambda=1}&=& 2\int_{r_{0}}^{r_{e}}\left(1-\frac{r _{0} ^{2}}{r^{2}}\right)^{-\frac{1}{2}}\left[\frac{2M}{r}+\frac{S}{r}+\frac{Mr _{0} }{r(r+r _{0} )}+\frac{Sr_{0}}{2r(r+r_{0})}\right]dr\nonumber\\
 &+&2 \int_{r_{0}}^{r_{s}}\left(1-\frac{r _{0} ^{2}}{r^{2}}\right)^{-\frac{1}{2}}\left[\frac{2M}{r}+\frac{S}{r}+\frac{Mr _{0} }{r(r+r _{0} )}+\frac{Sr_{0}}{2r(r+r_{0})}\right]dr.
\end{eqnarray}
Therefore, the required  delay in time  corresponding to $\lambda=1$ is

\begin{eqnarray}
\Delta T _{e},_{s}\mid_{\lambda=1}= 2\left(2M+S\right)\ln\left(\frac{\left(r_{e}+\sqrt{r_{e}^{2}-r_{0}^{2}}\right)\left(r_{s}+\sqrt{r_{s}^{2}-r_{0}^{2}}\right)}{r_{0}^{2}}\right)+
\left(2M+S\right)\left[\sqrt{\frac{r_{e}-r_{0}}{r_{e}+r_{0}}}+\sqrt{\frac{r_{s}-r_{0}}{r_{s}+r_{0}}}\right].
\end{eqnarray}

\subsection{Case $ \lambda$=2 }
\begin{eqnarray}
\Delta T _{e},_{s}\mid_{\lambda=2}&=& 2\int_{r_{0}}^{r_{e}}\left(1-\frac{r _{0} ^{2}}{r^{2}}\right)^{-\frac{1}{2}}\left[\frac{2M}{r}+\frac{S}{r^{2}}+\frac{Mr _{0} }{r(r+r _{0} )}+\frac{S(1+\frac{r_{0}}{r})}{2r(r+r_{0})}\right]dr\nonumber\\
 &+& 2\int_{r_{0}}^{r_{s}}\left(1-\frac{r _{0} ^{2}}{r^{2}}\right)^{-\frac{1}{2}}\left[\frac{2M}{r}+\frac{S}{r^{2}}+\frac{Mr _{0} }{r(r+r _{0} )}+\frac{S(1+\frac{r_{0}}{r})}{2r(r+r_{0})}\right]dr,
\end{eqnarray}
Therefore, the required  delay in time corresponding to $\lambda=2$ is
\begin{eqnarray}
\Delta T _{e},_{s}\mid_{\lambda=2}&=& 4M\ln\left(\frac{\left(r_{e}+\sqrt{r_{e}^{2}-r_{0}^{2}}\right)\left(r_{s}+\sqrt{r_{s}^{2}-r_{0}^{2}}\right)}{r_{0}^{2}}\right)+2M\left[\sqrt{\frac{r_{e}-r_{0}}{r_{e}+r_{0}}}+\sqrt{\frac{r_{s}-r_{0}}{r_{s}+r_{0}}}\right]\nonumber\\
&+& \frac{3S}{r_{0}}\left[\tan^{-1}\left(\frac{\sqrt{r_{e}^{2}-r_{0}^{2}}}{r_{0}}\right)+\tan^{-1}\left(\frac{\sqrt{r_{s}^{2}-r_{0}^{2}}}{r_{0}}\right)\right].
\end{eqnarray}
\subsection{Case $\lambda$=3}
\begin{eqnarray}
\Delta T _{e},_{s}\mid_{\lambda=3}&=& 2\int_{r_{0}}^{r_{e}}\left(1-\frac{r _{0} ^{2}}{r^{2}}\right)^{-\frac{1}{2}}\left[\frac{2M}{r}+\frac{S}{r^{3}}+\frac{Mr _{0} }{r(r+r _{0} )}+\frac{S\left(1+\frac{r_{0}}{r}+(\frac{r_{0}}{r})^{2}\right)}{2r(r+r_{0})r_{0}}\right]dr\nonumber\\
 &+& 2\int_{r_{0}}^{r_{s}}\left(1-\frac{r _{0} ^{2}}{r^{2}}\right)^{-\frac{1}{2}}\left[\frac{2M}{r}+\frac{S}{r^{3}}+\frac{Mr _{0} }{r(r+r _{0} )}+\frac{S\left(1+\frac{r_{0}}{r}+(\frac{r_{0}}{r})^{2}\right)}{2r(r+r_{0})r_{0}}\right]dr.
\end{eqnarray}
Therefore, the required delay in time corresponding to $\lambda=3$ is
\begin{eqnarray}
\Delta T _{e},_{s}\mid_{\lambda=3}&=& 4M\ln\left(\frac{\left(r_{e}+\sqrt{r_{e}^{2}-r_{0}^{2}}\right)\left(r_{s}+\sqrt{r_{s}^{2}-r_{0}^{2}}\right)}{r_{0}^{2}}\right)+2M\left[\sqrt{\frac{r_{e}-r_{0}}{r_{e}+r_{0}}}+\sqrt{\frac{r_{s}-r_{0}}{r_{s}+r_{0}}}\right]\nonumber\\
&+& 2S\left[\frac{\sqrt{r_{e}^{2}-r_{0}^{2}}}{r_{e}r_{0}^{2}}+\frac{\sqrt{r_{s}^{2}-r_{0}^{2}}}{r_{s}r_{0}^{2}}\right]+\frac{S}{r_{0}^{2}}\left[\sqrt{\frac{r_{e}-r_{0}}{r_{e}+r_{0}}}\left(\frac{r_{0}+2r_{e}}{r_{e}}\right)\right]+\frac{S}{r_{0}^{2}}\left[\sqrt{\frac{r_{s}-r_{0}}{r_{s}+r_{0}}}\left(\frac{r_{0}+2r_{s}}{r_{s}}\right)\right].
\end{eqnarray}
\subsection{Case $\lambda$=4}
\begin{equation*}
\Delta T _{e},_{s}\mid_{\lambda=4}= 2\int_{r_{0}}^{r_{e}}\left(1-\frac{r _{0} ^{2}}{r^{2}}\right)^{-\frac{1}{2}}\left[\frac{2M}{r}+\frac{S}{r^{4}}+\frac{Mr _{0} }{r(r+r _{0} )}+\frac{S\left(1+\frac{r_{0}}{r}+(\frac{r_{0}}{r})^{2}+(\frac{r_{0}}{r})^{3}\right)}{2r(r+r_{0})r_{0}^{2}}\right]dr
\end{equation*}
\begin{equation}
 + 2\int_{r_{0}}^{r_{s}}\left(1-\frac{r _{0} ^{2}}{r^{2}}\right)^{-\frac{1}{2}}\left[\frac{2M}{r}+\frac{S}{r^{4}}+\frac{Mr _{0} }{r(r+r _{0} )}+\frac{S\left(1+\frac{r_{0}}{r}+(\frac{r_{0}}{r})^{2}+(\frac{r_{0}}{r})^{3}\right)}{2r(r+r_{0})r_{0}^{2}}\right]dr.
\end{equation}

Therefore, the required  delay in time corresponding to $\lambda=4$ is
\begin{eqnarray}
\Delta T _{e},_{s}\mid_{\lambda=4}&=& 4M\ln\left(\frac{\left(r_{e}+\sqrt{r_{e}^{2}-r_{0}^{2}}\right)\left(r_{s}+\sqrt{r_{s}^{2}-r_{0}^{2}}\right)}{r_{0}^{2}}\right)+2M\left[\sqrt{\frac{r_{e}-r_{0}}{r_{e}+r_{0}}}+\sqrt{\frac{r_{s}-r_{0}}{r_{s}+r_{0}}}\right]\nonumber\\
&+& \frac{5S}{2r_{0}^{3}}\left[\tan^{-1}\left(\frac{\sqrt{r_{e}^{2}-r_{0}^{2}}}{r_{0}}\right)+\tan^{-1}\left(\frac{\sqrt{r_{s}^{2}-r_{0}^{2}}}{r_{0}}\right)\right]+ \frac{3S}{r_{0}^{2}}\left[\frac{\sqrt{r_{e}^{2}-r_{0}^{2}}}{r_{e}^{2}}+\frac{\sqrt{r_{s}^{2}-r_{0}^{2}}}{r_{s}^{2}}\right].
\end{eqnarray}

\section{Light deflection by charged and massless Wormholes in massive gravity }

Let us set the mass to zero i.e. $M=0$ and introduce the following coordinate transformation $r^\lambda=S+v^2$ into the metric (6), in that case we find the wormhole solution given by the Einstein-Rosen (ER) bridge form
\begin{equation}
ds^2=-\frac{v^2}{v^2+S}dt^2+\frac{4 dv^2}{\lambda^2 (S+v^2)^{\frac{\lambda-2}{\lambda}}}+(S+v^2)^{2/\lambda}d\Omega^2_{2}.
\end{equation}

The throat of the wormhole is located $v = 0$, with radius $R_{thro.}=S^{\frac{1}{\lambda}}$. This metric represents a massless wormhole with scalar charge $S$, and as far as we know this is a new metric. One can check by setting $\lambda=2$ and $S=-Q^2$ the above metric takes the form of usual charged ER wormhole.  From now on, we shall consider $v = r$, in this way from the metric (104)  the Lagrangian yields
\begin{eqnarray}
2\,\mathcal{L}=\left(\frac{r(s)^2}{r(s)^2+S}\right)\dot{t}^2(s)+\frac{4\dot{r}^2(s)}{\lambda^2\left (S+r(s)^2\right)^{(\lambda-2)/\lambda}}+(r^2(s)+S)^{2/\lambda}\left(\dot{\theta}^2(s)+\sin^2\theta \dot{\varphi}^2   (s)\right)
\end{eqnarray}

Going through same procedure and introducing a new variable $r=1/u$ as in the black hole case, we find the following equation
\begin{equation}
\frac{4}{\lambda^2 u^4 \mathcal{Z}}\left(\frac{du}{d\varphi}\right)^2-\left(S+\frac{1}{u}\right)^{(4+\lambda)/\lambda}\frac{u^2}{b^2}+\left(S+\frac{1}{u}\right)^{2/\lambda}=0
\end{equation}
where
\begin{equation}
\mathcal{Z}=\frac{S}{\left(S+\frac{1}{u}\right)^{2/\lambda}}+\frac{1}{\left(S+\frac{1}{u}\right)^{2/\lambda} u^2}.
\end{equation}

On the other hand the wormhole optical metric reads
\begin{equation}
dt^2=\frac{4 (S+r^2)^{2/\lambda} dr^2}{\lambda^2 r^2 }+\frac{(S+r^2)^{(2+\lambda)/\lambda}}{r^2}d\varphi^2,
\end{equation}
with 
\begin{equation}
dr^{\star}=\frac{2 (S+r^2)^{1/\lambda} dr}{\lambda r },\,\,\, f(r^{\star})=\frac{(S+r^2)^{(2+\lambda)/2\lambda}}{r}.
\end{equation}

The Gaussian optical curvature is found to be
\begin{equation}
K=-\frac{S\lambda \left[(\lambda+1) r^2+\frac{S\lambda}{2}  \right]}{2 (r^2+S)^{2 (\lambda+1)/\lambda}}.
\end{equation}

We shall consider the deflection angle by the spacetime metric (104) in terms of the GB method.

\subsection{Case $\lambda=1$}

The Gaussian optical curvature from Eq. (109) in the case when $\lambda=1$ reads
\begin{equation}
K_{\lambda=1} \simeq -\frac{S}{r^6}
\end{equation}

Substituting this result into Eq. \eqref{GBT2} generates the value of the deflection angle in terms of the integral
\begin{eqnarray}
\hat{\alpha}_{\lambda=1}^{GB}&=&-\int\limits_{0}^{\pi}\int\limits_{r_{\gamma}}^{\infty}\left(-\frac{S}{r^6} \right)\sqrt{\det g^{op}}\,dr^{\star}d\varphi.
\end{eqnarray}

In order to evaluate the above integral we need to find the equation for the light ray which can be found from Eq. (106) which yields 
\begin{equation}
\frac{4}{u^2(S+u)^2}\left(\frac{du}{d\varphi}\right)^2-\left(S+\frac{1}{u}\right)^{5}\frac{u^2}{b^2}+\left(S+\frac{1}{u}\right)^{2}=0
\end{equation}

If we linearize Eq. (113) around $S$, and then consider the equation which corresponds to straight line approximation we are left with the following equation
\begin{equation}
2 \left(\frac{du}{d\varphi}\right)^2+2 u \frac{d^2u}{d\varphi^2}+u^2=0.
\end{equation}

Solving this differential equation and using the condition $u(0)=0$ and $u(\pi/2)=1/b$ we find
\begin{equation}
u=\frac{\sqrt{\sin\varphi}}{b}.
\end{equation}

Finally the light ray equation in terms of the old coordinate gives
\begin{equation}
r_{\gamma}=\frac{b}{\sqrt{\sin\varphi}}.
\end{equation}

The deflection angle is found to be
\begin{equation}
\hat{\alpha}_{\lambda=1}^{GB} \simeq -\int\limits_{0}^{\pi}\int\limits_{\frac{b}{\sqrt{\sin\varphi}}}^{\infty}\left(-\frac{S}{r^6} \right)\sqrt{\det g^{op}}\,dr^{\star}d\varphi = \frac{2S}{b^2}.
\end{equation}

\subsection{$\lambda=2$}
In this case when $\lambda=2$ the Gaussian optical curvature yields
\begin{equation}
K_{\lambda=2} \simeq -\frac{3S}{r^4}
\end{equation}

We Substitute this equation in the deflection angle led to the following integral
\begin{eqnarray}
\hat{\alpha}_{\lambda=2}^{GB}&=&-\int\limits_{0}^{\pi}\int\limits_{r_{\gamma}}^{\infty}\left(-\frac{3S}{r^4} \right)\sqrt{\det g^{op}}\,dr^{\star}d\varphi.
\end{eqnarray}

Considering a series expansion around $S$ in Eq. (106) and then take only the  straight line approximation led to the following differential equation
\begin{equation}
\frac{d^2 u}{d\varphi^2}+u=0.
\end{equation}

Solving this equation we find the light ray equation
\begin{equation}
r_{\gamma}=\frac{b}{\sin\varphi}.
\end{equation}

Using the above result for the deflection angle we find
\begin{equation}
\hat{\alpha}_{\lambda=2}^{GB}\simeq -\int\limits_{0}^{\pi}\int\limits_{\frac{b}{\sin\varphi}}^{\infty}\left(-\frac{3S}{r^4} \right)\sqrt{\det g^{op}}\,\mathrm{d}r^{\star}\mathrm{d}\varphi = \frac{3 \pi S}{4 b^2}.
\end{equation}

\subsection{$\lambda=3$}
The Gaussian optical curvature in the case when $\lambda=3$ is found to be
\begin{equation}
K_{\lambda=3} \simeq -\frac{6 S}{r^{4}}
\end{equation}

From the GBT we find
\begin{eqnarray}
\hat{\alpha}_{\lambda=3}^{GB}&=&-\int\limits_{0}^{\pi}\int\limits_{r_{\gamma}}^{\infty}\left(-\frac{6 S}{r^{4}} \right)\sqrt{\det g^{op}}\,dr^{\star}d\varphi.
\end{eqnarray}

On the other hand the light ray equation in this case reduces to a nonlinear differential equation.  However we can approximate this equation from Eq. (106) as follows
\begin{equation}
\frac{4}{9}\frac{d^2 u}{d \varphi^2}+u=0.
\end{equation}

Solving this equation one finds
\begin{equation}
r_{\gamma}=\frac{b}{\sqrt{2}\sin(\frac{3\varphi}{2})}
\end{equation}

Using the above result for the deflection angle we find
\begin{equation}
\hat{\alpha}_{\lambda=3}^{GB}=-\int\limits_{0}^{\pi}\int\limits_{\frac{b}{\sqrt{2}\sin(\frac{3\varphi}{2})}}^{\infty}\left(-\frac{6 S}{r^{4}} \right)\sqrt{\det g^{op}}\,dr^{\star}d\varphi = \frac{15 \,S \,\Gamma\left(\frac{5}{6}\right)\Gamma\left(\frac{2}{3}\right)\sqrt{3} \,\,2^{1/3}}{16\, b^{8/3} \,\sqrt{\pi}}
\end{equation}

\subsection{$\lambda=4$}
We start by calculating first the Gaussian optical curvature when $\lambda=4$ to find
\begin{equation}
K_{\lambda=4} \simeq -\frac{10S}{r^3}.
\end{equation}

This result with the help of GBT gives
\begin{eqnarray}
\hat{\alpha}_{\lambda=4}^{GB}&=&-\int\limits_{0}^{\pi}\int\limits_{r_{\gamma}}^{\infty}\left(-\frac{10S}{r^3} \right)\sqrt{\det g^{op}}\,dr^{\star}d\varphi.
\end{eqnarray}

From Eq. (106) we find as follows
\begin{equation}
\frac{d^3 u}{d\varphi^3}+4 \frac{d u}{d\varphi}=0,
\end{equation}
with the following equation for the light ray
\begin{equation}
r_{\gamma}=\frac{2b}{1-\cos(2\varphi)}.
\end{equation}

Using the above result for the deflection angle we find
\begin{equation}
\hat{\alpha}_{\lambda=4}^{GB}\simeq -\int\limits_{0}^{\pi}\int\limits_{\frac{2b}{1-\cos(2\varphi)}}^{\infty}\left(-\frac{10S}{r^3} \right)\sqrt{\det g^{op}}\,dr^{\star}d\varphi = \frac{15 S \pi}{16 b^2}.
\end{equation}

Thus we have shown that the deflection angle increases with the increase of the parameter $\lambda$ for a constant value of the scalar charge $S$. It is a straightforward calculation to show and check these results in terms of the geodesic approach.

\begin{figure}[h!]
\center
\includegraphics[width=0.45\textwidth]{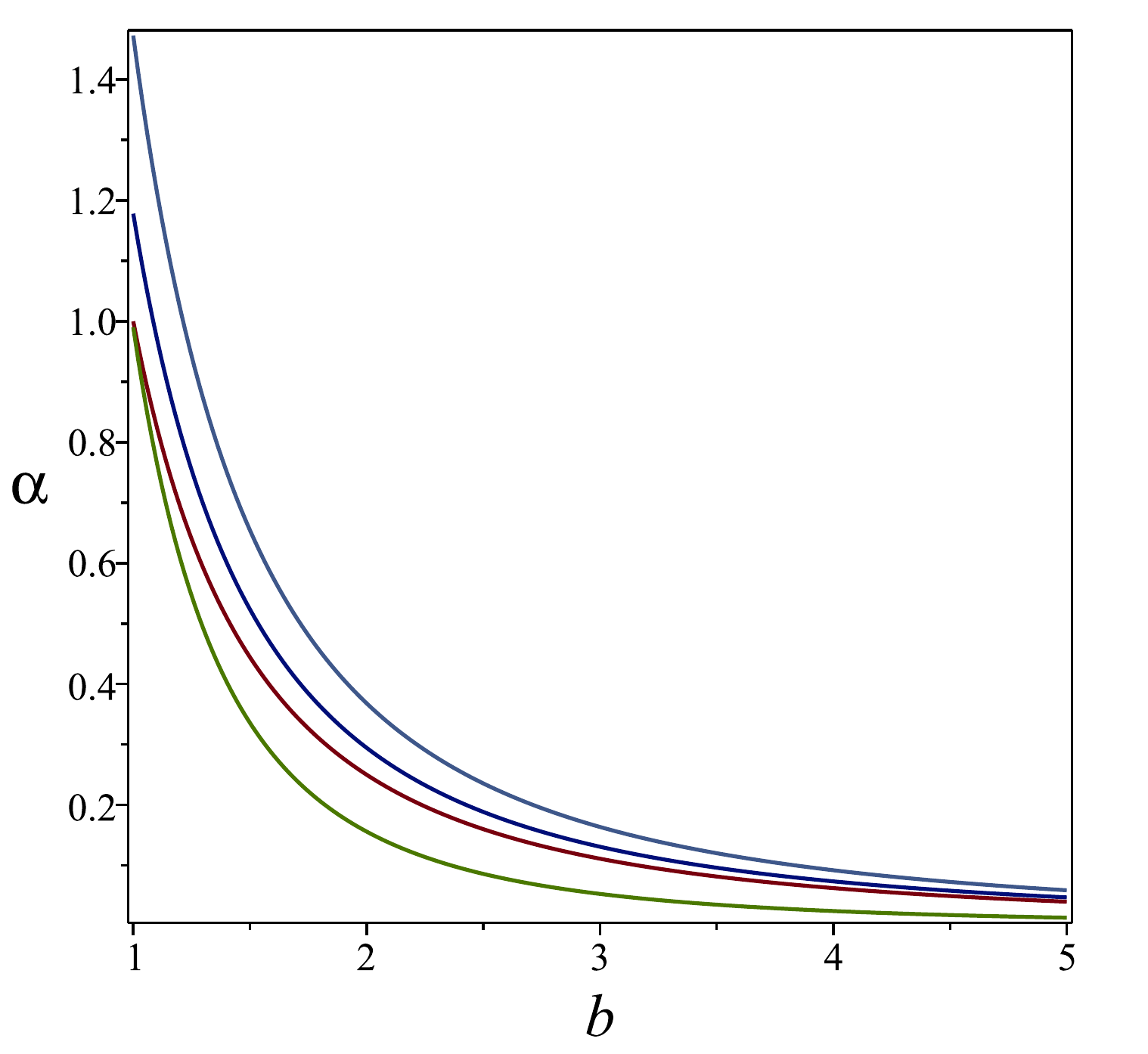}
\caption{\small \textit{We plot the deflection angle as a function of the impact factor $b$. We have chosen $S=0.5$. We see that with the increase of $\lambda$ the deflection angle actually increases.}}
\end{figure}
 
\newpage
\section{Time delay due to massless wormhole in massive gravity}

Here, we focus to estimate the time delay due to the massless wormholes in the massive gravity. Using the same technique as above, we calculate the delay in time for the  cases corresponding to the values of $\lambda = 1, 2, 3$  and 4 respectively.

\subsection{Case $\lambda =1$}
Here, we find the time delay as

\begin{eqnarray}
\Delta T|_{\lambda=1}&=&(2S) ln\left(\frac{\left[(S+v_e^2)+\sqrt{(v_e^2-v_0^2)(v_e^2+v_0^2+2S)}\right]\left[(S+v_e^2)+\sqrt{(v_s^2-v_0^2)(v_s^2+v_0^2+2S)}\right]}{(S+v_0^2)^2}\right)\nonumber\\
&\times& S\left[\sqrt{\frac{v_e^2-v_0^2}{v_e^2+v_0^2+2S}}+\sqrt{\frac{v_s^2-v_0^2}{v_s^2+v_0^2+2S}}\right].
\end{eqnarray}

\subsection{Case $\lambda =2$}
Here $S+v^2=r^2$, hence $S+v_e^2=r_e^2$ and $S+v_0^2=r_0^2$. In this case, we obtain  the time delay as
\begin{equation}
\Delta T|_{\lambda=2}=\frac{3S}{\sqrt{S+v_0^2}}\left[tan^{-1}\left(\frac{\sqrt{(v_e^2-v_0^2)}}{\sqrt{S+v_0^2}}\right)+tan^{-1}\left(\frac{\sqrt{(v_s^2-v_0^2)}}{\sqrt{S+v_0^2}}\right)\right].
\end{equation}

\subsection{Case $\lambda =3$}

Corresponding the value of $\lambda =3$, time delay is found as 

\begin{eqnarray}
\Delta T|_{\lambda=3} = 2S\left[\frac{\sqrt{(S+v_e^2)^{\frac{2}{3}}-(S+v_0^2)^{\frac{2}{3}}}}{(S+v_e^2)^{\frac{1}{3}}(S+v_0^2)^{\frac{2}{3}}}+\frac{\sqrt{(S+v_s^2)^{\frac{2}{3}}-(S+v_0^2)^{\frac{2}{3}}}}{(S+v_s^2)^{\frac{1}{3}}(S+v_0^2)^{\frac{2}{3}}}\right]\nonumber\\
+\frac{S}{(S+v_0^2)^{\frac{2}{3}}}\left[\left(\sqrt{\frac{(S+v_e^2)^{\frac{1}{3}}-(S+v_0^2)^{\frac{1}{3}}}{(S+v_e^2)^{\frac{1}{3}}+(S+v_0^2)^{\frac{1}{3}}}}\right)
\left(\frac{(S+v_0^2)^{\frac{1}{3}}+2(S+v_e^2)^{\frac{1}{3}}}{(S+v_e^2)^{\frac{1}{3}}}\right)\right]\nonumber\\
+\frac{S}{(S+v_0^2)^{\frac{2}{3}}}\left[\left(\sqrt{\frac{(S+v_s^2)^{\frac{1}{3}}-(S+v_0^2)^{\frac{1}{3}}}{(S+v_s^2)^{\frac{1}{3}}+(S+v_0^2)^{\frac{1}{3}}}}\right)
\left(\frac{(S+v_0^2)^{\frac{1}{3}}+2(S+v_s^2)^{\frac{1}{3}}}{(S+v_s^2)^{\frac{1}{3}}}\right)\right].
\end{eqnarray}

\subsection{Case $\lambda =4$}

In this case  we calculate the time delay as

\begin{eqnarray}
\Delta T|_{\lambda=4}=\frac{5S}{2(S+v_0^2)^{\frac{3}{4}}}\left[tan^{-1}\left(\frac{\sqrt{(S+v_e^2)^{\frac{1}{2}}-(S+v_0^2)^{\frac{1}{2}}}}{(S+v_0^2)^{\frac{1}{4}}}\right)
+tan^{-1}\left(\frac{\sqrt{(S+v_s^2)^{\frac{1}{2}}-(S+v_0^2)^{\frac{1}{2}}}}{(S+v_0^2)^{\frac{1}{4}}}\right)\right]\nonumber\\
 +\frac{3S}{(S+v_0^2)^{\frac{1}{2}}}\left[\left(\frac{\sqrt{(S+v_e^2)^{\frac{1}{2}}-(S+v_0^2)^{\frac{1}{2}}}}{(S+v_e^2)^{\frac{1}{2}}}\right)+\left(\frac{\sqrt{(S+v_s^2)^{\frac{1}{2}}-(S+v_0^2)^{\frac{1}{2}}}}{(S+v_s^2)^{\frac{1}{2}}}\right)\right].
\end{eqnarray}

\section{Conclusions}

In this paper we have studied the weak gravitational lensing for a black hole and wormhole in massive gravity. The black hole solution is governed by a parameter $\lambda$ dependent further on the mass $M$ and scalar charge $S$. 
In the case of  vanishing $S$, the results of the standard Schwarzschild geometry are recovered. By deforming the black hole solution in terms of the following coordinate transformation $r^{\lambda}=S+v^2$ we constructed a wormhole solution of ER type bridge which is regular in the interval  $-\infty <v < \infty$.  The deflection angle is then computed for four different values of the parameter $\lambda$. The extension of this work via Gauss--Bonnet theorem is nontrivial. First we derive a result showing how the Gaussian optical curvature and deflection angle is to be computed. The analysis is aided through the use Taylor series expansions. The time delay function is also established and computed for each of the four cases of $\lambda$ of interest in this investigation. Graphical plots indicate that for a fixed value of the mass and positive scalar charge, the deflection angle decreases with increasing $\lambda$, while for negative scalar charge, the deflection angle increases with an increase in $\lambda$. Whereas in the wormhole case we found that the deflection angle increases with the increase of the parameter $\lambda$ for a constant value of the scalar charge $S$, provided $S > 0$. 
\\
\\

\textbf{Acknowledgments}:   AB and FR are thankful to the authority of
Inter-University Centre for Astronomy and Astrophysics, Pune,
India for providing research facilities.  FR  and  NS  are  also  grateful  to  DST-SERB
(Grant No.:  EMR/2016/000193) and CSIR (Grant No.:
 09/096(0863)/2016-EMR-I), Govt.  of India for financial
support respectively. AB wishes to thank the University of KwaZulu-Natal (ACRU) for financial support.

\end{document}